\documentclass[twocolumn,showpacs,preprintnumbers,amsmath,amssymb,superscriptaddress]{revtex4}

\usepackage[dvipdfmx]{graphicx}
\usepackage{graphicx}
\usepackage{dcolumn}
\usepackage{color}
\usepackage{amsmath, amssymb}
\usepackage{bm}

\begin{document}

\preprint{APS/Physical Review A}

\title{Semiclassical quench dynamics of Bose gases in optical lattices}

\author{Kazuma Nagao}
\email{kazuma.nagao@yukawa.kyoto-u.ac.jp}
\affiliation{%
Yukawa Institute for Theoretical Physics, Kyoto University, Kitashirakawa Oiwakecho, Sakyo-ku, Kyoto 606-8502, Japan
}%
\author{Masaya Kunimi}%
\affiliation{%
Yukawa Institute for Theoretical Physics, Kyoto University, Kitashirakawa Oiwakecho, Sakyo-ku, Kyoto 606-8502, Japan
}%
\author{Yosuke Takasu}%
\affiliation{%
Department of Physics, Kyoto University, Kitashirakawa Oiwakecho, Sakyo-ku, Kyoto 606-8502, Japan
}%
\author{Yoshiro Takahashi}%
\affiliation{%
Department of Physics, Kyoto University, Kitashirakawa Oiwakecho, Sakyo-ku, Kyoto 606-8502, Japan
}%
\author{Ippei Danshita}%
\affiliation{%
Yukawa Institute for Theoretical Physics, Kyoto University, Kitashirakawa Oiwakecho, Sakyo-ku, Kyoto 606-8502, Japan
}%
\affiliation{%
Department of Physics, Kindai University, 3-4-1 Kowakae, Higashi-Osaka, Osaka 577-8502, Japan
}%

\date{\today}

\begin{abstract}

We analyze the time evolution of the Bose-Hubbard model after a sudden quantum quench to a weakly interacting regime. Specifically, motivated by a recent experiment at Kyoto University, we numerically simulate redistribution of the kinetic and onsite-interaction energies at an early time, which was observed in non-equilibrium dynamics of ultracold Bose gases in a cubic optical lattice starting with a singly-occupied Mott-insulator state. In order to compute the short-time dynamics corresponding to the experimental situation, we apply the truncated-Wigner approximation (TWA) to the Bose-Hubbard model on a cubic lattice. We show that our semiclassical approach quantitatively reproduces the fast redistribution dynamics. We further analyze spatial spreading of density-density correlations at equal time in the Bose-Hubbard model on a square lattice with a large filling factor. When the system is initially prepared in a coherent state, we find that a propagation velocity of the correlation wave packet in the correlation function strongly depends on the final interaction strength, and it is bounded by twice the maximum group velocity of the elementary excitations. In contrast, when the system is initially in a Mott-insulator state, the propagation velocity of the wave packet is approximately independent of the final interaction strength.

\end{abstract}

\pacs{Valid PACS appear here}
\maketitle

\section{\label{sec:level1}Introduction}\label{Sec: Intro}

High controllability and cleanness of ultracold-gas systems allow us to utilize them as analog quantum simulators for quantum many-body systems~\cite{Bloch_2012,Gross_2017,Hofstetter_2018}. Indeed, the performance of quantum simulators built with ultracold gases in optical lattices has been validated by the quantitative agreement between experiment and exact numerical simulation~\cite{Trotzky_2010,Endres_2011,Trotzky_2012,Cheneau_2012,Mazurenko_2017}. An interesting target of such quantum simulators is quantum dynamics far from equilibrium~\cite{Polkovnikov_2011,Eisert_2015}, which is in general impossible to simulate with currently available numerical methods on classical computers due to the exponential growth of the Hilbert-space dimension with system size and the minus-sign problem in quantum Monte Carlo simulations. Direct comparison with numerical simulations by the time-dependent density matrix renormalization group (t-DMRG) at one dimension (1D) has demonstrated that the quantum simulator can provide accurate data even in a long-time region, where t-DMRG fails~\cite{Trotzky_2012}.

Among diverse quantum many-body dynamics, particular attention has been devoted to quantum quench dynamics, which arises after a sudden and substantial change of parameters in the Hamiltonian~\cite{Trotzky_2012,Cheneau_2012,Greiner_2002b,Altman_2002,Tuchman_2006,Kollath_2007,Dziarmaga_2012,Meinert-13, Polkovnikov_2011,Eisert_2015}. In recent years, some experimental groups have explored far-from-equilibrium dynamics of high-dimensional Bose-Hubbard systems quenched from typical quantum states~\cite{Braun_2015,Braun_thesis,Asaka_2016,Takasu_2018b}. Specifically, some of the current authors and their collaborators observed the redistribution dynamics of kinetic and onsite-interaction energies of Bose gases in a cubic optical lattice after a rapid quench of the lattice depth from a Mott insulator state~\cite{Asaka_2016,Takasu_2018b}. An immediate usage of such quantum simulation results is to examine or develop numerical methods for computing quantum many-body dynamics by taking them as an accurate reference. Nevertheless, any quantitative approach that can recover the experimental results at three dimensions (3D) has not been established thus far.

In this paper, aiming to simulate the energy-redistribution dynamics quantitatively, we adopt a semiclassical approximation formulated by a phase-space representation of quantum systems, namely, the truncated-Wigner approximation (TWA) \cite{Hillery_1984,Steel_1998,Blakie_2008,Polkovnikov_2010}. According to the framework of TWA for the Bose-Hubbard model \cite{Polkovnikov_2010}, one can represent a time-dependent quantum average of physical observables as a semiclassical form in terms of deterministic trajectories of a discrete Gross--Pitaevskii equation, where the initial classical fields are weighed with the Wigner's quasi-probability distribution corresponding to an initial quantum state. This approximation allows one to obtain quantitative descriptions of short-time dynamics of the quantum averages even for macroscopic quantum systems, to which exact-diagonalization methods are inaccessible. 

In the past two decades, TWA or related semiclassical frameworks were widely used to explore non-equilibrium phenomena of isolated Bose gases trapped by optical lattices \cite{Polkovnikov_2002,Polkovnikov_2003a,Polkovnikov_2003b,Polkovnikov_2004,Tuchman_2006,Polkovnikov_2010,Mathey_2009,Mathey_2010,Mathey_2017,Dziarmaga_2012,Landea_2015,Fujimoto_2018,Kunimi_2018}, quantum spin systems \cite{Polkovnikov_2010,Davidson_2015,Schachenmayer_2015a,Schachenmayer_2015b,Wurtz_2018}, open quantum systems \cite{Gardiner_2004,Rey_2014,Kordas_2015,Johnson_2017,Vicentini_2018}, spin-boson models \cite{Altland_2009,Orioli_2017,Raventos_2018}, and interacting fermions \cite{Davidson_2017,Scaffidi_2017,Schmitt_2018}. In earlier literatures on interacting bosons in optical lattices \cite{Polkovnikov_2002,Polkovnikov_2003a,Tuchman_2006,Dziarmaga_2012}, it was argued that in a weakly interacting regime the semiclassical approach can be used to describe the time evolution induced by a sudden quench from a Mott-insulator state. However, the application of such semiclassical approaches to the 3D case at unit filling, which is the situation realized in the experiment \cite{Asaka_2016,Takasu_2018b}, has not been demonstrated in practice. We apply a TWA technique, which was previously used to study 1D Bose-Hubbard dynamics at a large-filling factor~\cite{Polkovnikov_2002,Polkovnikov_2003a,Tuchman_2006}, to the case of the 3D Bose-Hubbard model initially prepared in a singly-occupied Mott-insulator state. By computing the time evolution of the kinetic and interaction energies inside a weakly interacting regime, we will show that the results obtained by our semiclassical approach agree well with those observed in the experiment \cite{Asaka_2016,Takasu_2018b}.

As a further application of the developed method, we study dynamics of non-local spreading of density-density spatial correlations after a sudden quench in the 2D Bose-Hubbard model. Correlation spreading in 1D ultracold gases has been discussed~\cite{Cheneau_2012,Lauchli_2008,Barmettler_2012} in the context of the Lieb-Robinson bound of non-relativistic quantum many-body systems~\cite{Lieb_1972}. Both of the experiment and theory showed that in the strongly interacting region the propagation speed of the correlation spreading is bounded by the maximum group velocity of particle-hole excitations. More recently, correlation-spreading dynamics at 2D have been analyzed~\cite{Takasu_2018b,Carleo_2014,Krutitsky_2014}, whereas their quantitative properties are, however, less understood compared with the 1D case. In this paper, within the semiclassical regime of the 2D Bose-Hubbard model, we compute the time evolution of a density-density correlation function at equal time by starting with either a coherent state or a Mott-insulator state. We find that when the system is initially prepared in a coherent state, a mean propagation velocity of a wave packet in the correlation function strongly depends on the final interaction. In contrast, when the system is initially in a Mott-insulator state, a wave packet in the correlation function propagates with a nearly constant velocity with respect to the final interaction.

The remaining part of this paper is organized as follows: In Sec.~\ref{Sec: TWA}, we introduce the Bose-Hubbard Hamiltonian, which effectively describes ultracold bosonic atoms tightly trapped by an optical-lattice potential, and explain the TWA method for this model. In Sec.~\ref{Sec: 3d}, we discuss an application of the TWA to the quench experiment at 3D~\cite{Takasu_2018b}. In Sec.~\ref{Sec: 2d}, we analyze spreading dynamics of spatial density-density correlations in a weakly interacting regime of the 2D Bose-Hubbard model after a sudden quench from either a coherent state or a Mott-insulator state. In Sec.~\ref{Sec: conclusions}, conclusions of this article are summarized.

\section{Truncated-Wigner approximation for the Bose-Hubbard Hamiltonian}\label{Sec: TWA}

In this paper, we investigate non-equilibrium dynamics of ultracold bosonic atoms trapped in an optical lattice. Supposing that the lattice depth is sufficiently deep, the system is effectively described by the single-band Bose-Hubbard model~\cite{Fisher_1989,Jaksch_1998}, 
\begin{align}
{\hat H} = -J\sum_{\langle jk \rangle}({\hat a}^{\dagger}_{j}{\hat a}_{k}+{\rm h.c.}) + \frac{U}{2}\sum_{j}{\hat a}^{\dagger}_{j}{\hat a}^{\dagger}_{j}{\hat a}_{j}{\hat a}_{j}. \label{eq: bhh}
\end{align}
Here ${\hat a}_{j}$ and ${\hat a}^{\dagger}_{j}$ are the annihilation and creation Bose operators at each site with an index $j$. The bracket symbol $\langle jk \rangle$ denotes nearest-neighbor pairs of the sites. The two energy scales $J$ and $U(>0)$ characterize the tunneling between nearest-neighbor sites and the onsite repulsive interaction. When the mean density of atoms per site (filling factor) is integer, the model (\ref{eq: bhh}) exhibits a quantum phase transition between the superfluid and Mott-insulator states \cite{Fisher_1989,Greiner_2002a}. The accurate values of the transition points at unit filling have been obtained with quasi-exact numerical methods as $(U/J)_{c} = 3.367$ (1D) \cite{Kuhner_2000}, $(U/J)_{c}=16.74$ (2D) \cite{Sansone_2008}, and $(U/J)_{c}=29.34$ (3D) \cite{Sansone_2007}, respectively. In the following discussions, we write $M$ and $N_{\rm tot}$ as the total numbers of lattice points and atoms.

Let us briefly review the TWA method applied to the Bose-Hubbard model. In terms of the phase space method defined in a $2M$-dimensional phase space of a complex-valued vector $\vec{\alpha}=(\alpha_1,\alpha_2,\cdots,\alpha_{M})$, the time evolution of lattice bosons can be described by a quasi-probability distribution, i.e. the Wigner function $W(\vec{\alpha},\vec{\alpha}^*,t)$, which is equivalent to the Wigner--Weyl transform of the density operator ${\hat \rho}(t)$ of the system \cite{Hillery_1984,Gardiner_2004,Blakie_2008,Polkovnikov_2010}. For the Bose-Hubbard model (\ref{eq: bhh}), the equation of motion of $W(\vec{\alpha},\vec{\alpha}^*,t)$ is given by
\begin{align}
i\hbar \frac{\partial}{\partial t}W(\vec{\alpha},\vec{\alpha}^*,t) = 2H_W(\vec{\alpha},\vec{\alpha}^*) {\rm sinh}\left(\frac{\Lambda_{c}}{2}\right) W(\vec{\alpha},\vec{\alpha}^*,t), \label{eq: time_wigner}
\end{align}
where $H_W = ({\hat H})_{W}$ is the Wigner--Weyl transform of Eq.~(\ref{eq: bhh}) \cite{Blakie_2008,Polkovnikov_2010}. The symbol $\Lambda_{c}$ represents the symplectic operator working on $c$-number functions defined in the phase space, and its explicit form reads 
\begin{align}
\Lambda_{c}=\sum_{j}\left[ \frac{\overleftarrow \partial}{\partial \alpha_{j}}\frac{\overrightarrow \partial}{\partial \alpha^*_{j}} - \frac{\overleftarrow \partial}{\partial \alpha^*_{j}}\frac{\overrightarrow \partial}{\partial \alpha_{j}} \right].
\end{align}
With use of the Wigner function, the time-dependent quantum average of an operator $\hat \Omega$, defined by $\langle {\hat \Omega}(t) \rangle \equiv {\rm Tr}[{\hat \rho}_0 {\hat \Omega}(t)]$, can be expressed as a phase-space averaged form \cite{Blakie_2008,Polkovnikov_2010}
\begin{align}
\langle {\hat \Omega}(t) \rangle = \int d\vec{\alpha}d\vec{\alpha}^{*} W(\vec{\alpha},\vec{\alpha}^*,t)\Omega_W(\vec{\alpha},\vec{\alpha}^*),
\end{align}
where $d\vec{\alpha}d\vec{\alpha}^{*} = \pi^{-M}\prod_{j}d{\rm Re}[\alpha_{j}]d{\rm Im}[\alpha_{j}]$.

According to the previous literatures \cite{Hillery_1984,Polkovnikov_2010}, the TWA is derived in a semiclassical expansion of the right-hand side of Eq.~(\ref{eq: time_wigner}) in the symplectic operator $\Lambda_{c}$. If one eliminates higher-order terms of order $O(\Lambda_{c}^3)$ from the expansion series, then the time evolution of the Wigner function is effectively generated by the classical Liouville equation $i \hbar \partial{W}/\partial t \approx \{H_{W},W\}_{\rm P.B.}$. Here the bracket $\{\cdot,\cdot\}_{\rm P.B.}$ denotes the Poisson bracket defined in the phase space. In this approximation, the Wigner function is conserved along characteristic trajectories, which are solutions of the discrete Gross--Pitaevskii equation $i\hbar \partial{\alpha}_{{\rm cl},j}/\partial t=\partial H_W/\partial \alpha_{{\rm cl},j}^{*}$ \cite{Polkovnikov_2010}. This statement is nothing but the Liouville theorem in the classical statistical mechanics \cite{Hillery_1984}. Using the theorem, we find that the quantum average of ${\hat \Omega}(t)$ can be reduced to a semiclassical form \cite{Blakie_2008,Polkovnikov_2010}
\begin{align}
\langle {\hat \Omega}(t) \rangle \approx \int d\vec{\alpha}_0d\vec{\alpha}^{*}_0 W(\vec{\alpha}_0,\vec{\alpha}_0^*) \Omega_{W}[\vec{\alpha}_{\rm cl}(t),\vec{\alpha}^{*}_{\rm cl}(t)], \label{eq: twa}
\end{align}
where $\vec{\alpha}_{\rm cl}(t)$ is a solution of the Gross--Pitaevskii equation for an initial classical field $\vec{\alpha}_0$. The initial classical fields distribute over the phase space according to the Wigner function of the initial quantum state. Here we note that the classical field scales with a square root of the filling factor $\bar{n} = N_{\rm tot}/M$, so that the expansion in $\Lambda_{c}$ is characterized by the inverse of ${\bar n}$ \cite{Polkovnikov_2010}.

The semiclassical approximation used in Eq.~(\ref{eq: twa}) yields a quantum-fluctuation correction to a mean-field solution of dynamics within the lowest order \cite{Polkovnikov_2003b}. In a weakly fluctuating regime, where an interaction parameter $\lambda \equiv U{\bar n}/J$ is far from the quantum phase transition point $\lambda_c$, i.e., $\lambda \ll \lambda_{c}$, the TWA quantitatively describes time evolution of the system until the time $t$ approaches a characteristic timescale $t_c$ \cite{Polkovnikov_2002,Polkovnikov_2003a,Polkovnikov_2010}. When $\lambda$ is close to the critical value $\lambda_{c}$, the semiclassical treatment breaks down at short time due to the strong fluctuations. Since $\lambda_c \propto {\bar n}^2$, larger $\bar{n}$ and/or smaller $U/J$ means larger $t_c$ \cite{Polkovnikov_2002,Polkovnikov_2003a}. Especially at $U/J = 0$ or $\bar{n}=\infty$, the semiclassical approximation becomes exact. 

In typical experiments including the one in Refs.~\cite{Asaka_2016,Takasu_2018b}, ${\bar n}$ is tuned to unity and $\lambda$ is $O(1)$. If one computes time evolution of the 1D Bose-Hubbard model with ${\bar n}=1$ and $\lambda \sim 1$ within the TWA, it fails in much shorter time than $O(\hbar/J)$ because of rather small $\lambda_c(=3.367)$. In contrast, for the 3D case with the same parameters, it is expected that the TWA is able to simulate the dynamics up to $t \sim \hbar/J$, because $\lambda$ of $O(1)$ is sufficiently far from $\lambda_c=29.34$. As we will see in Sec.~\ref{Sec: 3d}, the TWA can reproduce characteristic early-time dynamics observed in the experiment \cite{Takasu_2018b} until $t \sim \hbar/J$.

\section{Redistribution dynamics of the kinetic and interaction energies}\label{Sec: 3d}

In this section, we apply the formalism of TWA for simulating the non-equilibrium dynamics of Bose gases in a cubic optical lattice observed in the experiment \cite{Asaka_2016,Takasu_2018b}. We numerically compute the time evolution of the kinetic and interaction energies after a sudden quench from a singly-occupied Mott-insulator state into a weakly interacting regime. 

\subsection{Experimental setup}

In the experiment of Refs.~\cite{Asaka_2016,Takasu_2018b}, a gas of $^{174}{\rm Yb}$ atoms (bosons) is confined in a cubic optical lattice with lattice spacing $d_{\rm lat}=266$ nm. The typical energy scale of this system is given by the recoil energy $E_{\rm R}/\hbar = 2\pi \times4021.18$ Hz. The experimental protocol for studying quantum quench dynamics is summarized as follows:

(i) One slowly ramps up the optical lattice depth $V_0$ up to $V_0=15E_{\rm R}$, at which $U/J = 99.4$, in order to prepare a singly-occupied Mott insulator.

(ii) One abruptly ramps down the lattice depth from $V_0 = 15 E_{\rm R}$ to $V_0 = 5 E_{\rm R}$ in the ramp-down time $t_{\rm f} = 0.1$ ms in order to prepare a state far from equilibrium. At the final depth, implying that $U/J = 3.41$, the ground state is deeply in the superfluid regime~\cite{Sansone_2007}.

(iii) After the ramp-down process, one measures the time evolution of ensemble averages of the kinetic energy ${\hat K} = -J\sum_{\langle j k \rangle}({\hat a}^{\dagger}_{j}{\hat a}_{k}+{\rm h.c.})$ and the onsite-interaction energy ${\hat O} = \frac{U}{2}\sum_{j}{\hat a}^{\dagger}_{j}{\hat a}^{\dagger}_{j}{\hat a}_{j}{\hat a}_{j}$. The kinetic and interaction energies are extracted from the time-of-flight imaging and the high-resolution atom-number-projection spectroscopy, respectively~\cite{Takasu_2018a}. 
  
It is worth noting that although there is a parabolic trapping potential in the real experiment \cite{Takasu_2018b}, we neglect it in our TWA calculations for the following reason. At the initial Mott-insulator state, the particle density of the system is almost uniform in space so that the initial quantum state is well approximated as a direct-product wave function, which is spatially uniform and composed of a local Fock state (see Sec.~\ref{Subsec: TWA}). The trapping potential gives no noticeable effect on the quench dynamics within the time window $t \lesssim \hbar/J$, in which the experiment was performed, because the trap frequency is much smaller than $J/\hbar$.

\subsection{Application of TWA to the quench experiment} \label{Subsec: TWA}

Within the framework of TWA, the time-dependent quantum mechanical average of the kinetic and interaction energies is approximated to a semiclassical form with the deterministic Gross--Pitaevskii trajectory $\vec{\alpha}_{\rm cl}(t)$
\begin{align}
\langle {\hat K}(t) \rangle &\approx \int d\vec{\alpha}_0d\vec{\alpha}^*_0W(\vec{\alpha}_0,\vec{\alpha}_0^*)K_{W}\left[ \vec{\alpha}_{\rm cl}(t),\vec{\alpha}^*_{\rm cl}(t) \right], \nonumber \\
\langle {\hat O}(t) \rangle &\approx \int d\vec{\alpha}_0d\vec{\alpha}^*_0W(\vec{\alpha}_0,\vec{\alpha}_0^*)O_{W}\left[ \vec{\alpha}_{\rm cl}(t),\vec{\alpha}^*_{\rm cl}(t) \right], \nonumber
\end{align}
where $K_W(\vec{\alpha},\vec{\alpha}^*)$ and $O_W(\vec{\alpha},\vec{\alpha}^*)$ are the Wigner-Weyl transforms of ${\hat K}$ and ${\hat O}$, respectively. The explicit forms of $K_W(\vec{\alpha},\vec{\alpha}^*)$ and $O_W(\vec{\alpha},\vec{\alpha}^*)$ can be derived by means of the Bopp-operator representation of the bosonic operators ${\hat a}_{j} \rightarrow \alpha_{j}+\frac{1}{2}\frac{\partial}{\partial \alpha_j^*}$ and ${\hat a}^\dagger_{j}\rightarrow \alpha^*_{j}-\frac{1}{2}\frac{\partial}{\partial \alpha_j}$ \cite{Polkovnikov_2010}.

The initial state before the quench in the experiment can be represented by a product-state wave function $|\Psi_{\rm ini}\rangle=\prod_{j}|{\bar n} \rangle_{j}$, where $|{\bar n} \rangle_{j}$ is a local Fock state characterized by ${\hat n}_{j}|{\bar n} \rangle_{j}={\bar n}|{\bar n} \rangle_{j}$. The corresponding Wigner function $W(\vec{\alpha},\vec{\alpha}^*)$ is given by a direct product of the local Wigner function of the Fock-state vector $|{\bar n}\rangle_{j}$ at each site, thus, it reads \cite{Gardiner_2002,Polkovnikov_2003a,Olsen_2004,Blakie_2008}
\begin{align}
W(\vec{\alpha},\vec{\alpha}^*) = \prod_{j}2e^{-2|\alpha_{j}|^2}(-1)^{{\bar n}}L_{{\bar n}}(4|\alpha_{j}|^2), \label{eq: Wigner_Mott}
\end{align}
where $L_{n}(x)=\sum_{r=0}^{n}(-1)^{r}\frac{n!}{(n-r)!(r!)^2}x^{r}$ is the Laguerre polynomial of order $n$. Here we parametrize the classical field as $\alpha_{j}=|\alpha_{j}|e^{i\varphi_j}$. This Wigner function is not positive definite along a direction of the amplitude degrees of freedom $|\alpha_j|$, except for a trivial case ${\bar n}=0$. The phase of the classical field $\varphi_j$ distributes uniformly in $[0,2\pi]$. The Wigner function has an explicit $U(1)$ symmetry reflecting the restored symmetry inside the Mott-insulator state. In fact, a general phase shift of the phase-space variables, $\alpha_{j} \rightarrow \alpha_{j}e^{i{\tilde \varphi}_{j}}$, does not change in the value of the Wigner function. 

The negativity of Eq.~(\ref{eq: Wigner_Mott}) makes it difficult to obtain converged results in numerically evaluating the phase-space integration weighted by the Wigner function. For this reason, in our numerical simulations, we adopt a Gaussian approximation for the exact Wigner function of a Fock state \cite{Gardiner_2002,Olsen_2004}. Repeating the discussions in the previous literatures \cite{Gardiner_2002,Olsen_2004}, the Gaussian-Wigner function corresponding to a Fock-state vector $|{\bar n}\rangle$ has a general form as
\begin{align}
W_{\rm g}(n) = \frac{1}{\sqrt{2\pi \sigma^2}}e^{-\frac{1}{2\sigma^2}(n-n_0)^2}, \label{eq: gauss}
\end{align}
where $n=|\alpha|^2$. The mean $n_0$ and covariance $\sigma$ are free parameters determined from the consistency that the Gaussian function should exactly recover the first and second order local moments of the density, i.e., $\langle {\hat n}_j \rangle$ and $\langle {\hat n}^2_j \rangle$. From direct calculations, we find that the optimal choice is $n_0={\bar n}+\frac{1}{2}$ and $\sigma = \frac{1}{2}$. It is worth noting that the (rescaled) higher-order moments ${\bar n}^{-m}\langle {\hat n}^{m}_{j} \rangle$ for $m>2$ computed by the Gaussian function agree with the exact ones up to $O({\bar n}^{-2})$ \cite{Olsen_2004}. While the normalized Gaussian function can give rise to an unphysical negative density, however, it does not affect the phase-space average itself because the probability, which corresponds to the Gaussian tail, is sufficiently small even at ${\bar n}=1$. In addition, a  similar Gaussian approximation is often used in literatures of TWA studies \cite{Gardiner_2002,Polkovnikov_2003a,Olsen_2004,Landea_2015,Davidson_2015,Davidson_2017,Wurtz_2018} and manifests its validity in the semiclassical descriptions of short time dynamics. 

\begin{figure}
\includegraphics[width=80mm]{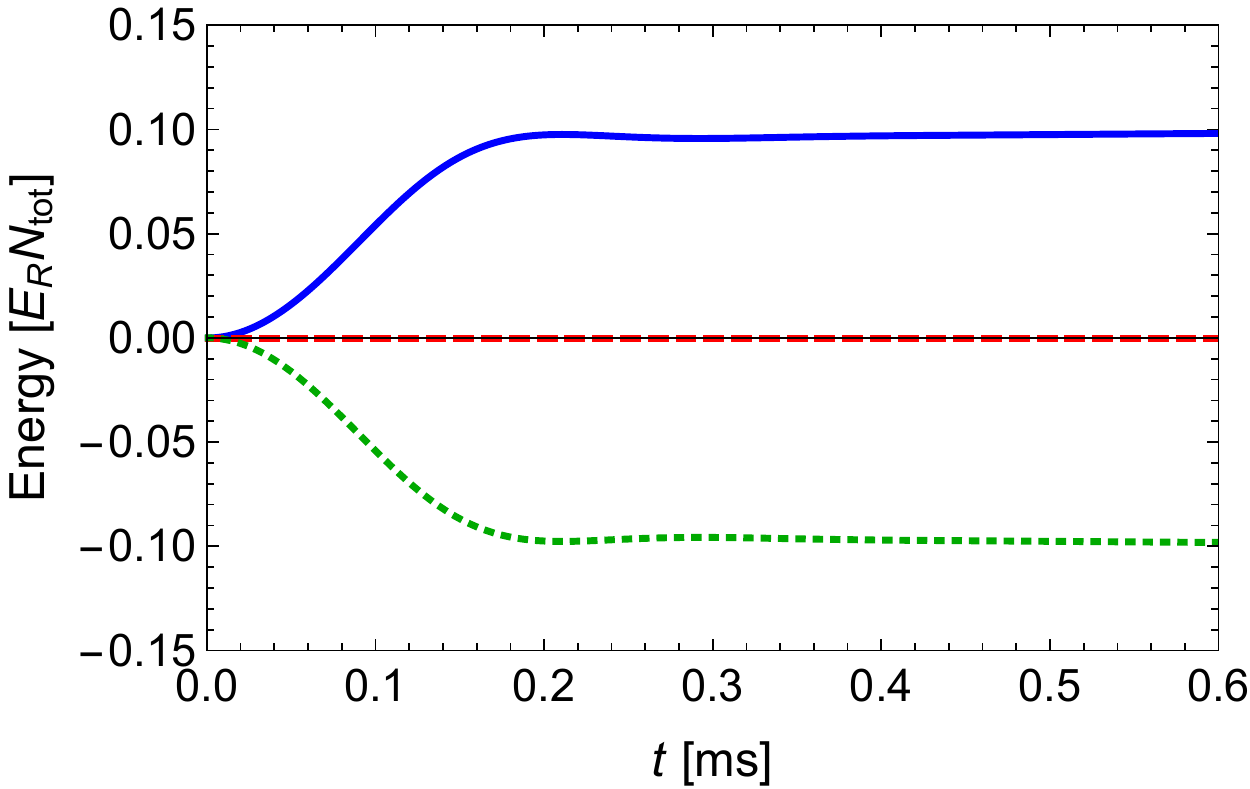}
\vspace{-3mm}
\caption{Semiclassical time evolution of the kinetic and onsite-interaction energies (green-dotted and blue-solid lines) after the sudden quench from the singly-occupied Mott-insulator state. We do not deal with finite-time effects of the ramp down process, i.e., $t_{\rm f}=0$. The red-dashed line represents the total sum of these energies. In the numerical simulation, we set $N_{\rm tot}=M=12^3$ and $\lambda_{\rm f}=3.41$. In the current setup, $0.6\;{\rm ms}\approx \hbar/J$, where $J$ corresponds to the final lattice depth. In this TWA simulation, we sampled $100000$ random initial conditions of the classical field according to Eq.~(\ref{eq: gauss}). Throughout this paper, we omit the standard error of the Monte-Carlo sampling because for each simulation the error is sufficiently small to be neglected.}
\label{figure001}
\end{figure}

Now we summarize what should be analyzed in the TWA: We solve time evolutions of a time-dependent Bose-Hubbard model ${\hat H}[\lambda(t)]$ by using the TWA. At $t=0$, the Hamiltonian has $\lambda = \lambda_{\rm i}=99.4$ corresponding to $V_0 = 15 E_{\rm R}$. In the ramp-down process, $\lambda(t)$ decreases with $V_{0}(t)$, which declines linearly. Recall that the duration of the ramp-down process is $t_{\rm f} = 0.1$ ms in the experiment. At $t=t_{\rm f}$, the lattice depth reaches $V_{0}=5E_{\rm R}$, which implies $\lambda = \lambda_{\rm f}=3.41$. At $t>t_{\rm f}$, the system evolves in time under the time-independent Hamiltonian ${\hat H}[\lambda_{\rm f}]$. The phase-space averaging with the Wigner function is evaluated by using the Monte-Carlo integration, where each initial configuration of the classical fields, $\vec{\alpha}_{\rm cl}(0)$, is randomly chosen from the Gaussian-Wigner function (\ref{eq: gauss}).

Before proceeding to a numerical simulation corresponding to the experimental setup, we discuss a simpler problem, i.e., an infinitesimal-time limit of the ramp-down process ($t_{\rm f}=0$). In this case, the Hamiltonian is always independent of time at $t>0$. Figure \ref{figure001} depicts a numerical simulation of the kinetic and interaction energies within TWA for $t_{\rm f}=0$, where we set $N_{\rm tot}=M=12^3$ and assume an open boundary condition. We clearly see that the semiclassical approach captures fast redistribution of the kinetic and interaction energies. The timescale of the redistribution is on the order of 0.1 ms and comparable to the experimental result. In addition, the sum of the energies, i.e., $E_{\rm tot}= \langle {\hat K}(t) \rangle + \langle {\hat O}(t) \rangle$ completely maintains its initial value because the Hamiltonian of the system is independent on time. We emphasize that the redistribution dynamics presented in Fig.~\ref{figure001} cannot be recreated by means of naive mean-field theories without fluctuations from a classical configuration, such as the Gross--Pitaevskii theory and the Gutzwiller variational method.

\subsection{TWA versus experimental results}\label{Subsec: twa_exp}

\begin{figure}
\includegraphics[width=85mm]{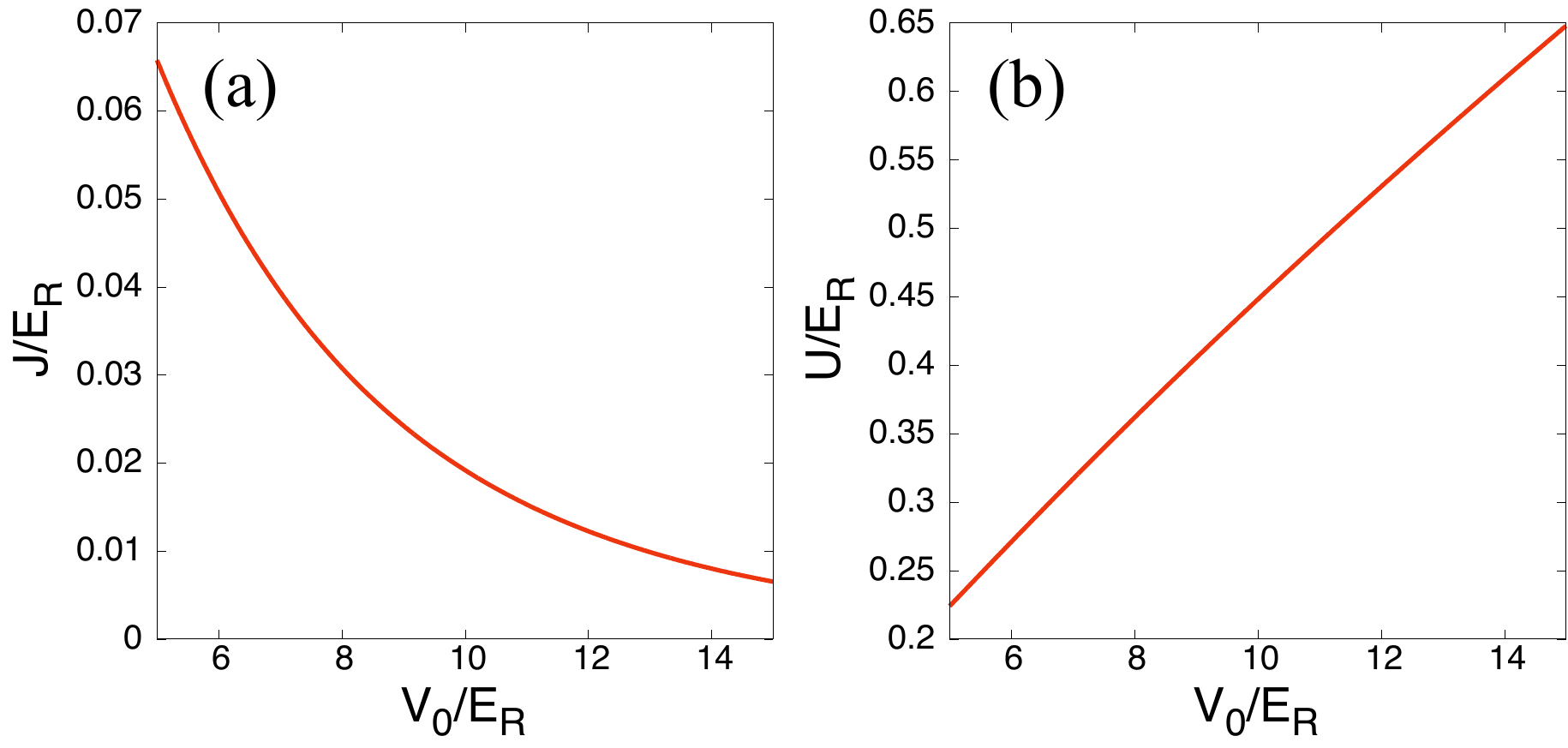}
\vspace{-4mm}
\caption{Dependence of (a) the hopping strength $J$ and (b) the onsite-interaction strength $U$ on the lattice depth $V_0$. These quantities are measured in the unit of the recoil energy $E_{\rm R}$.}
\label{figure002}
\end{figure}

Here we take into account the finite-time ramp-down process in $V_0(t)$. The hopping strength $J$ and the onsite-interaction strength $U$ vary with $V_0(t)$ as depicted in Fig.~\ref{figure002}. We note again that $V_0(t)$ linearly decreases in time from $V_0(0)=15E_{\rm R}$ to $V_0(t_{\rm f})=5E_{\rm R}$ where $t_{\rm f}=0.1$ [ms]. In this process, the system passes through the Mott-insulating and the quantum critical regimes where the quantitative validity of TWA is justified only in rather short time $t\ll O(\hbar/J)$. Nevertheless, our approach is expected to be able to explain the redistribution dynamics after the quench because the system actually leaves away from these regimes in the short time.

\begin{figure}
\includegraphics[width=80mm]{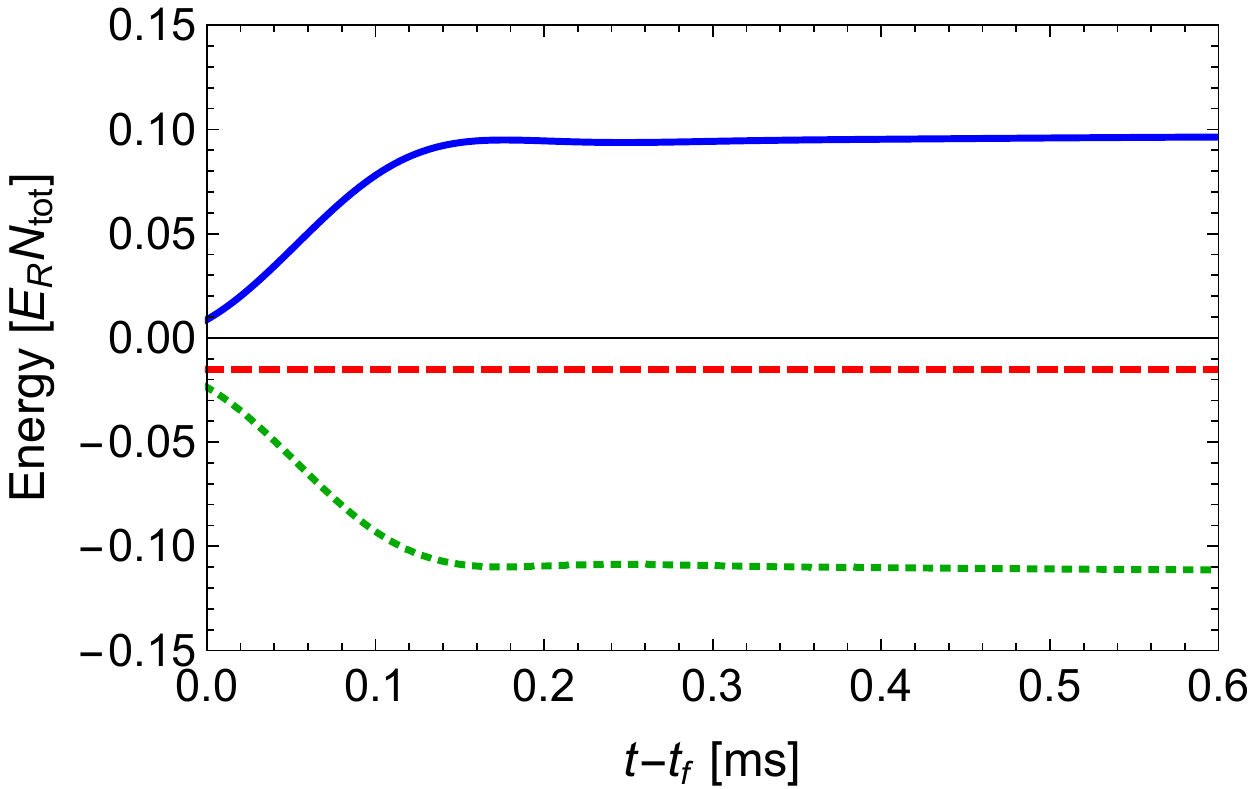}
\vspace{-3mm}
\caption{Semiclassical time evolution of the kinetic and onsite-interaction energies (green-dotted and blue-solid lines) including the ramp-down process from the singly-occupied Mott-insulator state. The red-dashed line represents the total sum of these energies. In the numerical simulation, we set $N_{\rm tot}=M=30^3$ and $\lambda_{\rm f}=3.41$. The horizontal axis starts from $t=t_{\rm f}$. In the TWA simulation, we sampled $10000$ initial conditions according to Eq.~(\ref{eq: gauss}).}
\label{figure003}
\end{figure}

In Fig.~\ref{figure003}, we show $\langle {\hat K}(t) \rangle$ and $\langle {\hat O}(t) \rangle$ including the ramp-down process. The numerical simulation is performed with an open boundary condition and at $M=N_{\rm tot}=30^3$, which is comparable to the size of the actual system. Compared with the previous calculation in Subsec.~\ref{Subsec: TWA}, the ramp-down process significantly modifies the value of each energy at $t=t_{\rm f}$. The total energy $E_{\rm tot}$ decreases from zero. In addition, the timescale for the saturation toward each quasi-steady value is slightly diminished. Due to such modifications, the semiclassical result including the ramp-down process agrees very well with the experimental one without any fitting parameter. A more direct comparison with the experimental result will be presented in Ref.~\cite{Takasu_2018b}.

We conclude this section by making comments on the limitation of our semiclassical approach to the experimental system at intermediate final interactions and at unit filling. Although the experiment is able to access a long-time regime $t \gg \hbar/J$, our approach is limited to simulate a short time dynamics up to $t \sim \hbar/J$. To develop an efficient tool which allows one to study the long-time dynamics, e.g., a relaxation dynamics toward a thermal equilibrium state, remains to be an open and challenging issue.

\section{Spatial-correlation spreading after a sudden quench}\label{Sec: 2d}

In Sec.~\ref{Sec: 3d}, we corroborated the quantitative validity of the TWA method for quantum quench dynamics of the 3D Bose-Hubbard model in a weakly interacting regime ($\lambda \gg \lambda_c$) starting with a Mott-insulator state. In this section, we next apply our approach to investigate time evolutions of spatial correlation spreading after sudden quenches in the 2D Bose-Hubbard model. Especially, we consider two different initial states, i.e., a coherent state, which is the ground state at $\lambda=0$, and a Mott-insulator state. We discuss their difference emerging in the resulting dynamics after a sudden quench into a weakly interacting regime. 

In order to characterize spatial-correlation spreading, we specifically deal with a density-density equal-time correlation function defined by
\begin{align}
C_{d}(t) = \frac{1}{M{\bar n}^2}\sum_{j} \langle {\hat n}_{j}(t){\hat n}_{j+d}(t) \rangle_{c}, \label{eq: density-density}
\end{align}
where $d=(d_x,d_y)$ is a 2D relative vector between two different sites. In the definition of the correlation function, $\langle \cdots \rangle_c$ denotes a connected correlation function, i.e., $\langle {\hat n}_{j}(t){\hat n}_{j+d}(t) \rangle_c = \langle {\hat n}_{j}(t){\hat n}_{j+d}(t) \rangle - \langle {\hat n}_{j}(t) \rangle \langle {\hat n}_{j+d}(t) \rangle$. Within TWA, the connected correlator is approximated to 
\begin{align}
\langle {\hat n}_{j}(t){\hat n}_{j+d}(t) \rangle_{c} \approx \overline{n_{W}^{(j)}(t)n_{W}^{(j+d)}(t)} - \overline{n_{W}^{(j)}(t)} \cdot \overline{n_{W}^{(j+d)}(t)},
\end{align}
where the overline in the right-hand side means the phase-space average by use of the Wigner function of initial quantum states. The $c$-number quantity $n_{W}^{(j)}$ represents the Wigner-Weyl transform of the local density ${\hat n}_j$, i.e., $n_{W}^{(j)} = |\alpha_{j}|^2-\frac{1}{2}$. In cold-atom experiments, the time evolution of the non-local density-density correlation is measurable by utilizing the quantum-gas microscope technique \cite{Cheneau_2012} or measuring spatial-noise correlations in a time-of-flight interference pattern of expanding gases \cite{Altman_2004}. 

\begin{figure}
\begin{center}
\includegraphics[width=80mm]{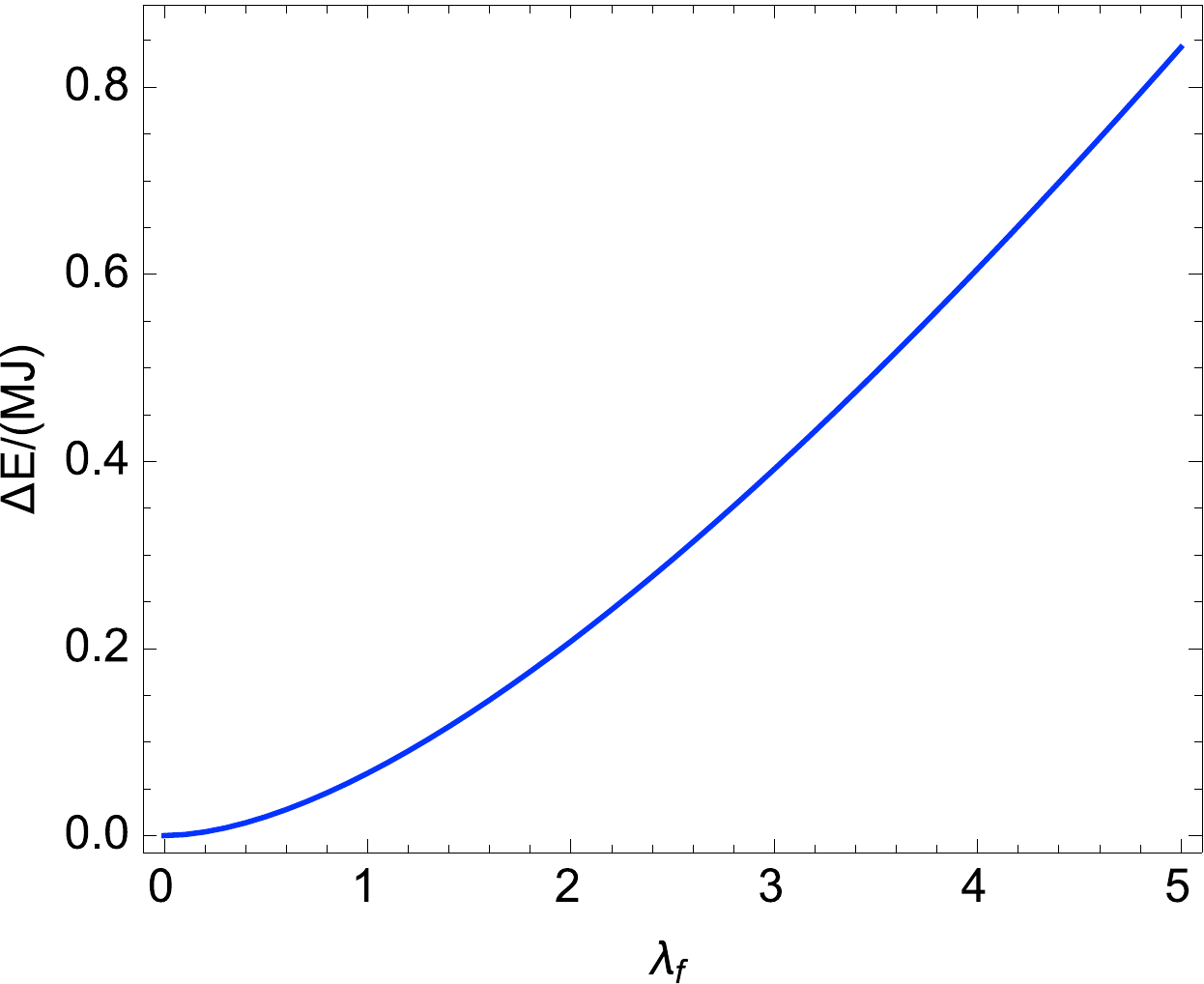}
\vspace{0.5mm}
\caption{Energy deviation per site $\Delta E/(MJ)$ of the coherent state from the ground state energy within the Bogoliubov approximation as a function of $\lambda_{\rm f}$, where $\bar{n}=10$. }
\label{figure004}
\end{center}
\end{figure}

\subsection{Sudden quench from a coherent state} \label{Subsec: SF}

We begin with analyzing density-density correlation spreading inside a superfluid regime assuming that the system is initially in a direct-product state composed of the local coherent states $|{\bar \alpha}\rangle_{j} = e^{{\bar \alpha}{\hat a}^{\dagger}_{j}-{\bar \alpha}^*{\hat a}_j}|0\rangle$:
\begin{align}
|\Psi_{\rm ini}\rangle = \prod_{j}| {\bar \alpha} \rangle_j. \label{eq: coherent}
\end{align}
Here, ${\bar \alpha} = \sqrt{\bar n}e^{i {\bar \varphi}}$ parametrizes each coherent-state vector. Calculating the Wigner-Weyl transform of this wave function (\ref{eq: coherent}), we can obtain the corresponding Wigner function as follows \cite{Polkovnikov_2010}: 
\begin{align}
W(\vec{\alpha},\vec{\alpha}^*) = \prod_{j}\left\{ 2e^{-2|\alpha_j - {\bar \alpha}|^2} \right\}. \label{eq: wigner_coherent}
\end{align}
This Wigner function can take on non-negative values for arbitrary $\alpha_j$, so that there is no difficulty in the Monte-Carlo sampling of the TWA. In the following discussions, we set ${\bar \varphi}=0$ for simplicity. 
 
\begin{figure}
\begin{center}
\includegraphics[width=70mm]{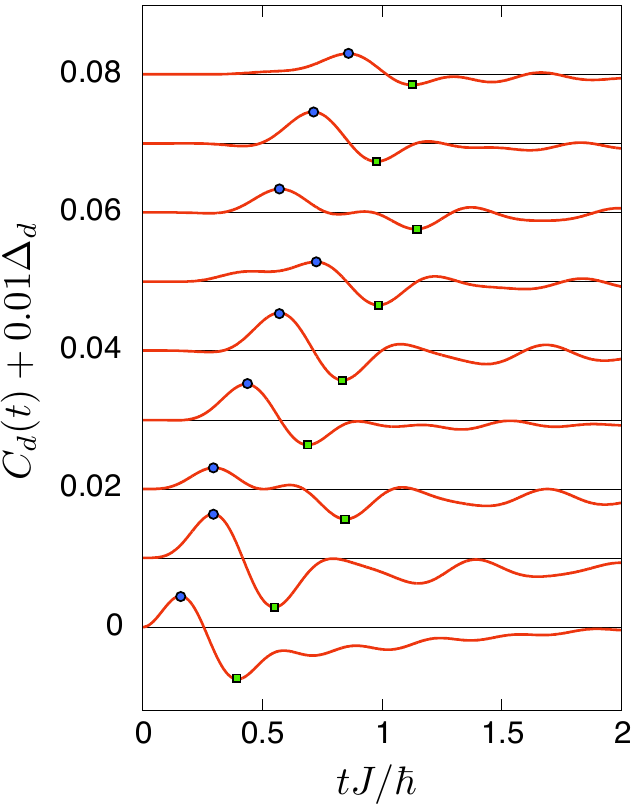}
\vspace{0.5mm}
\caption{Density-density correlation spreading after the sudden quench from the coherent state at $\lambda_{\rm f}=2$. The blue circle and green square indicate the maximum and minimum values of the correlation function within $tJ/\hbar \leq 3$. The relative vector $d=(d_x,d_y)$, Euclidean distance $d_{\rm E}$, and offset of correlation $\Delta_d$ take values of $(d_x,d_y;d_{\rm E};\Delta_d)=(0,1;1.00;0)$, $(1,1;1.41;1)$, $(0,2;2.00;2)$, $(1,2;2.24;3)$, $(2,2;2.83;4)$, $(0,3;3.00;5)$, $(1,3;3.16;6)$, $(2,3;3.617)$, $(0,4;4.00;8)$ from the bottom to top, respectively. In the TWA simulation, we sampled $40000$ initial conditions according to Eq.~(\ref{eq: wigner_coherent}).}
\label{figure005}
\end{center}
\end{figure}

\begin{figure*}
\begin{center}
\includegraphics[width=140mm]{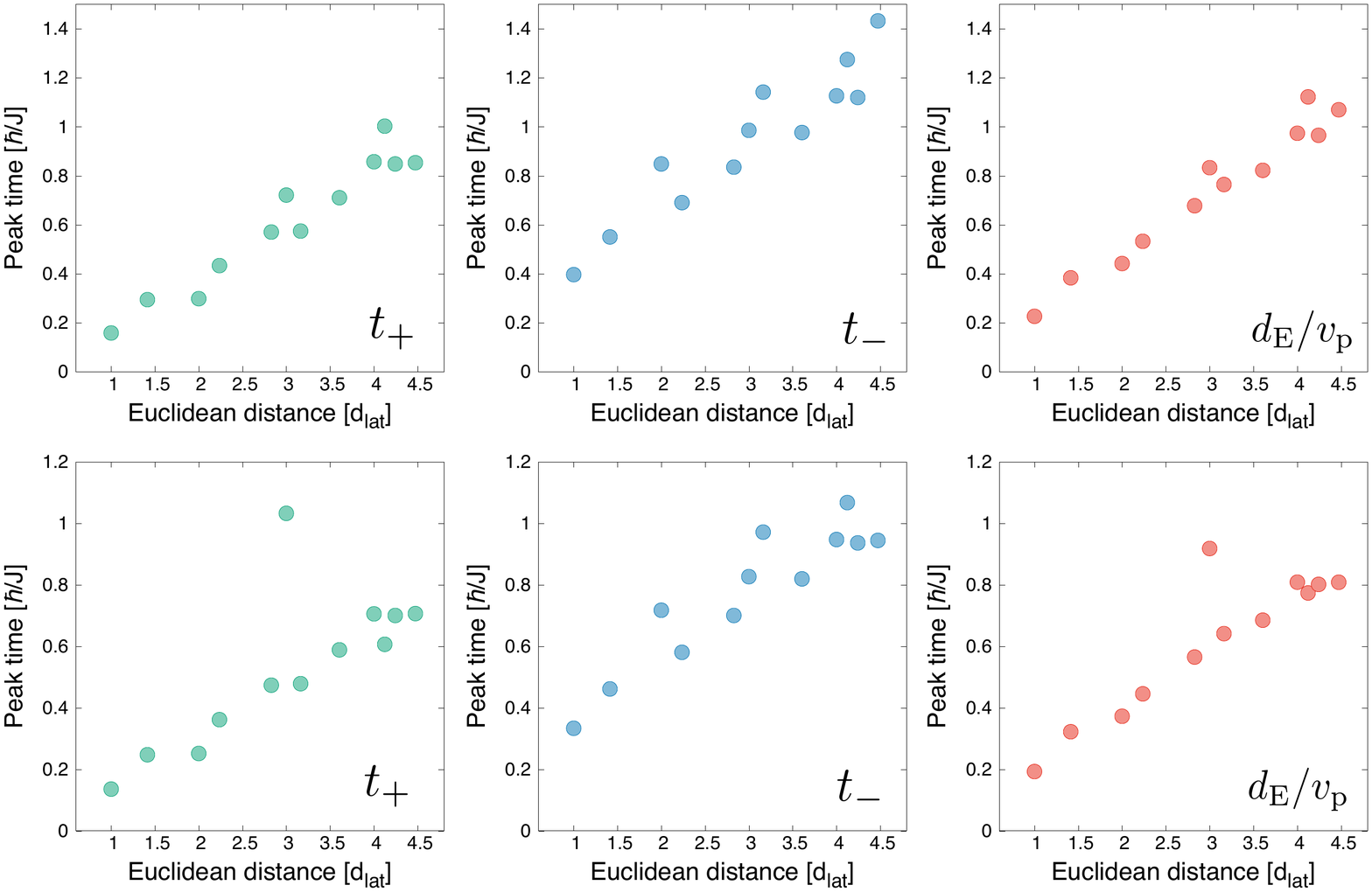}
\vspace{-10mm}
\caption{Maximum (left column), minimum (center column) and averaged (right column) peak times extracted from the TWA simulations. The vertical and horizontal axes express the peak time and Euclidean distance. The upper and lower rows correspond to $\lambda_{\rm f}=2$ and $\lambda_{\rm f}=4$, respectively. }
\label{figure006}
\end{center}
\end{figure*}

In order to keep the accuracy of TWA for a relatively long timescale, here we choose ${\bar n}=10$ in numerical simulations. In addition, we impose periodic boundary conditions on the system. Throughout this section, we suppose that the quench is abruptly done for an infinitesimal time, for simplicity.
 
Before proceeding to our main results, we calculate the energy deviation per site defined by
\begin{align} 
\frac{1}{M}\Delta E = \frac{1}{M}\left[\langle \Psi_{\rm ini}|{\hat H}_{\rm f} |\Psi_{\rm ini}\rangle - \langle {\hat H}_{\rm f} \rangle_{\rm g} \right],
\end{align}
where ${\hat H}_{\rm f}$ is the Hamiltonian at $\lambda=\lambda_{\rm f}$ and $\langle {\hat H}_{\rm f} \rangle_{\rm g}$ means the ground-state energy of ${\hat H}_{\rm f}$. We evaluate $\langle {\hat H}_{\rm f} \rangle_{\rm g}$ within the standard Bogoliubov approximation for the Bose-Hubbard model as follows: 
\begin{align}
\langle {\hat H}_{\rm f} \rangle_{\rm g} \approx M\left[ {\cal E}_{0} + \frac{1}{2M}\sum_{{\bf p}\neq 0}(E_{\bf p}-\hbar \omega_{\bf p}) \right],
\end{align}
where ${\bf p}=(p_x,p_y)$ is a momentum in the first Brillouin zone, ${\cal E}_{0}=-4J{\bar n}+U{\bar n}^2/2$, and $\hbar \omega_{\bf p}=U{\bar n}+4J\sum_{j=x,y}{\rm sin}^{2}[p_j d_{\rm lat}/(2\hbar)]$. In addition, $E_{\bf p}=\sqrt{(\hbar \omega_{\bf p})^2-({\bar n}U)^2}$ is the energy of the elementary excitations (For more details, see Ref.~\cite{Danshita_2009}). Figure \ref{figure004} shows $\Delta E/(MJ)$ of the coherent state as a function of $\lambda_{\rm f}$. Because $\Delta E/M$ is less than the typical energy scale $J$ over a wide range of $\lambda_{\rm f}$, the dynamics after the quench from the coherent state is dominated by the low-energy elementary excitations from the ground state, i.e., the Bogoliubov quasiparticles.
 
Figure \ref{figure005} monitors how density-density correlations propagate over the square lattice. In the numerical simulation, we set $\lambda_{\rm f}=2$ and $M=20^2$ at ${\bar n}=10$. In addition, we characterize the correlation spreading by means of the usual Euclidean distance defined by $d_{\rm E} \equiv ({d_x^2 + d_y^2})^{1/2}$. In the time evolution, we observe that a characteristic signal of correlation, i.e., a wave packet enveloping maximum (blue circle) and minimum (green square) peaks of a fine oscillation propagates over the square lattice in time. Such a fine oscillation can be interpreted as a quasi-coherent oscillation reflecting that a few elementary excitations are created by the quench.

To quantify the correlation spreading, we extract a propagation velocity of the wave packet from the numerical results in the following manner. Let us denote the peak times of the maximum and minimum values of the correlation as $t_{+}$ and $t_{-}$, which are represented by the blue circles and the green squares in Fig.~\ref{figure005}. For a given Euclidean distance $d_{\rm E}$, we can define a reasonable (instantaneous) propagation velocity $v_{\rm p}$ as a harmonic average of these peak times such that
\begin{align}
v_{\rm p} \equiv \frac{d_{\rm E}}{2}\left(\frac{1}{t_{+}}+\frac{1}{t_{-}}\right), \label{eq: def_group_vel}
\end{align}
where $d_{\rm E}/v_{\rm p}$ is regarded as an averaged peak time. In Fig.~\ref{figure006}, we indicate $t_+$, $t_-$, and $d_{\rm E}/v_{\rm p}$ for different relative distances at $\lambda_{\rm f}=2$ and $\lambda_{\rm f}=4$, respectively. It is found that the averaged peak time almost linearly increases with $d_{\rm E}$. A linear fitting of the averaged peak times gives a mean propagation velocity, ${\bar v}_{\rm p}$, of the wave packet. In order to compute ${\bar v}_{\rm p}$, we take into account early twelve peaks in a timescale of $t \sim \hbar/J$, which are found in $C_{d}(t)$ with $d_{\rm E}<5d_{\rm lat}$.

\begin{figure}
\begin{center}
\includegraphics[width=65mm]{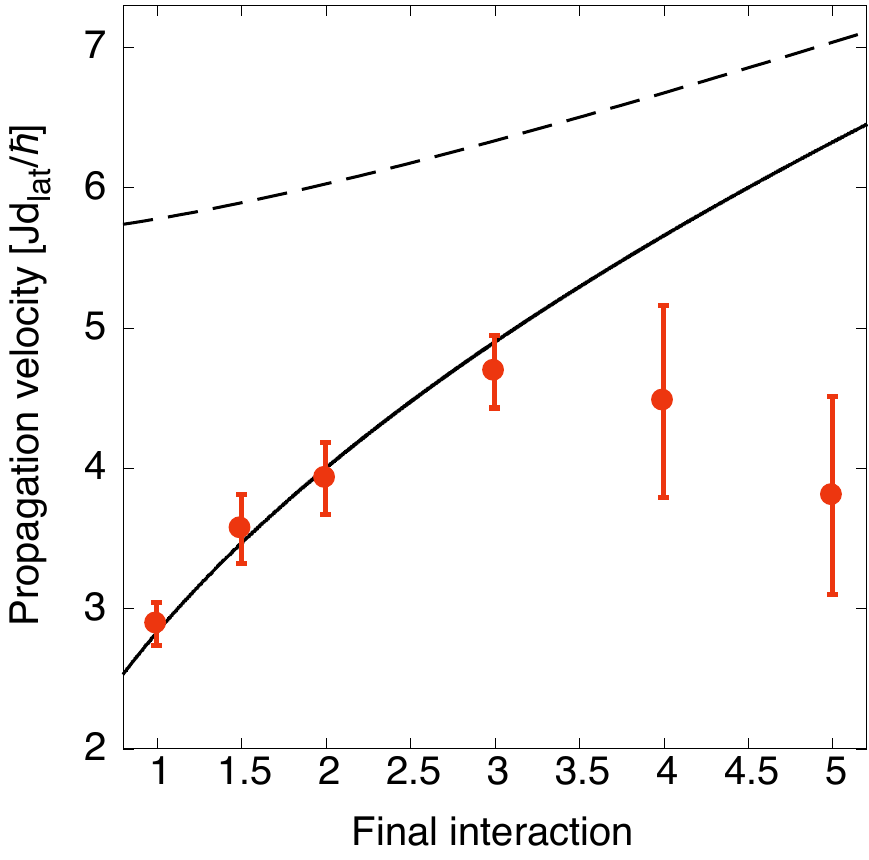}
\vspace{0.5mm}
\caption{Final interaction dependence of the mean propagation velocity ${\bar v}_{\rm g}$ (circle). The solid and dashed lines represent twice the sound ($2v_{\rm s}$) and maximum ($2v_{\rm m}$) velocities of the Bogoliubov excitation, respectively. The horizontal axis expresses the final interaction $\lambda_{\rm f}$. The vertical bar indicates the normal estimation error of the mean propagation velocity in the linear fitting (see also Fig.~\ref{figure006}).}
\label{figure007}
\end{center}
\end{figure}

Figure \ref{figure007} shows the mean propagation velocity ${\bar v}_{\rm p}$ as a function of the final interaction $\lambda_{\rm f}$. In the same figure, we also display twice the maximum and sound velocities of the Bogoliubov excitations, $2v_{\rm m}$ and $2v_{\rm s}$, which are expressed as
\begin{align}
v_{\rm m} &= \max_{\bf p}\left\{ \sqrt{\left(\frac{\partial E_{\bf p}}{\partial p_x}\right)^2+\left(\frac{\partial E_{\bf p}}{\partial p_y}\right)^2} \right\}, \nonumber \\
v_{\rm s}&= \lim_{{\bf p} \rightarrow {\bf 0}} \left\{ \sqrt{\left(\frac{\partial E_{\bf p}}{\partial p_x}\right)^2+\left(\frac{\partial E_{\bf p}}{\partial p_y}\right)^2} \right\}. \nonumber
\end{align}
Note that $v_{\rm m}$ coincides with $v_{\rm s}$ in the limit that $\lambda_{\rm f} \gg 1$. It is clearly observed in Fig.~\ref{figure007} that ${\bar v}_{\rm p}$ is bounded by twice the maximum velocity $2v_{\rm m}$ over a range of $\lambda_{\rm f} \in [1,5]$. This numerical result is consistent with the general one of the Lieb--Robinson bound \cite{Lieb_1972}. Furthermore, in the range of $1 \leq \lambda_{\rm f} \leq 3 $, the propagation velocity increases with $\lambda_{\rm f}$ in such a way that the points come close to $2v_{\rm s}$. This feature can be attributed to the fact that the quench actually creates some elementary excitations at $E_{\bf p} \ll J$, where the Bogoliubov excitations behave as phonons, because the energy deviation is relatively small (See Fig.~\ref{figure004}).

In contrast, in the range of $3 < \lambda_{\rm f} \leq 5$, the mean propagation velocity significantly deviates from $2v_{\rm s}$. In this regime, elementary excitations with $E_{\bf p}\sim J$ can be generated because the energy deviation per particle is comparable to $J$ as seen in Fig.~\ref{figure004}. In addition, the computed propagation velocity has a large estimation error of the linear fitting. The large error is actually due to an exceptional point in, e.g., $C_{d}(t)$ at $d_{\rm E}=3$ and $\lambda_{\rm f}=4$ (see Fig.~\ref{figure006}) that the maximum peak arises after the growth of the minimum one. Similar points also appear at $\lambda_{\rm f}=5$.

We conclude this subsection with comments on a previous study on similar quench dynamics of the 2D Bose-Hubbard model, which uses a time-dependent variational Monte-Carlo approach \cite{Carleo_2014}. In Ref.~\cite{Carleo_2014}, Carleo and coworkers calculated the density-density correlation function in a weakly-interacting regime starting from a superfluid ground state at ${\bar n}=1$. Figure 2(b) of Ref.~\cite{Carleo_2014} implies an unphysical result that the propagation velocity is much greater than twice the maximum one of the elementary excitation in the regime. While it seems to contradict the Lieb--Robinson bound, the crucial reason of such a fast propagation has not been mentioned in their paper. To characterize the wavefront motion of the correlation on the square lattice, in Ref.~\cite{Carleo_2014}, the propagation velocity was evaluated in terms of the Manhattan distance $d_{\rm M}\equiv|d_x|+|d_y|$ \cite{Carleo_2014,Krause_1986}. It is worth emphasizing that if we redefine ${\bar v}_{\rm p}$ by the Manhattan distance instead of the Euclidean one in our TWA results, it leads to a similar fast propagation as in Ref.~\cite{Carleo_2014}. Hence, we argue that the fast propagation beyond twice the maximum velocity seen in Ref.~\cite{Carleo_2014} is actually due to the unsuitable choice of the distance to define a propagation velocity.

\subsection{Quench from a Mott-insulator state across a quantum phase transition}

\begin{figure}
\begin{center}
\includegraphics[width=80mm]{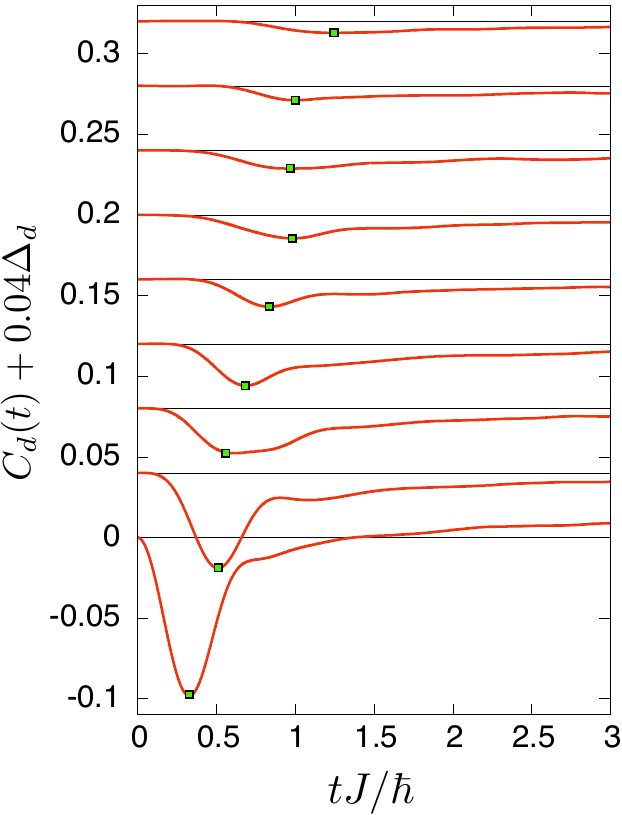}
\vspace{0.5mm}
\caption{Density-density correlation spreading after the sudden quench from the Mott-insulator state at $\lambda_{\rm f}=2$. The green square indicates the minimum peak of the correlation signal. The relative vector $d=(d_x,d_y)$, Euclidean distance $d_{\rm E}$, and offset of correlation $\Delta_d$ take values of $(d_x,d_y;d_{\rm E};\Delta_d)=(0,1;1.00;0)$, $(1,1;1.41;1)$, $(0,2;2.00;2)$, $(1,2;2.24;3)$, $(2,2;2.83;4)$, $(0,3;3.00;5)$, $(1,3;3.16;6)$, $(2,3;3.617)$, $(0,4;4.00;8)$ from the bottom to top, respectively. In the TWA simulation, we sampled $10000$ initial conditions according to Eq.~(\ref{eq: gauss}).}
\label{figure008}
\end{center}
\end{figure}

We now discuss a sudden quench from a Mott-insulator state and keep track of density-density correlation spreading, which occurs inside a weakly interacting regime. Note that the initial state corresponds to the ground state of the system at $\lambda=\infty$.

\begin{figure}
\begin{center}
\includegraphics[width=87mm]{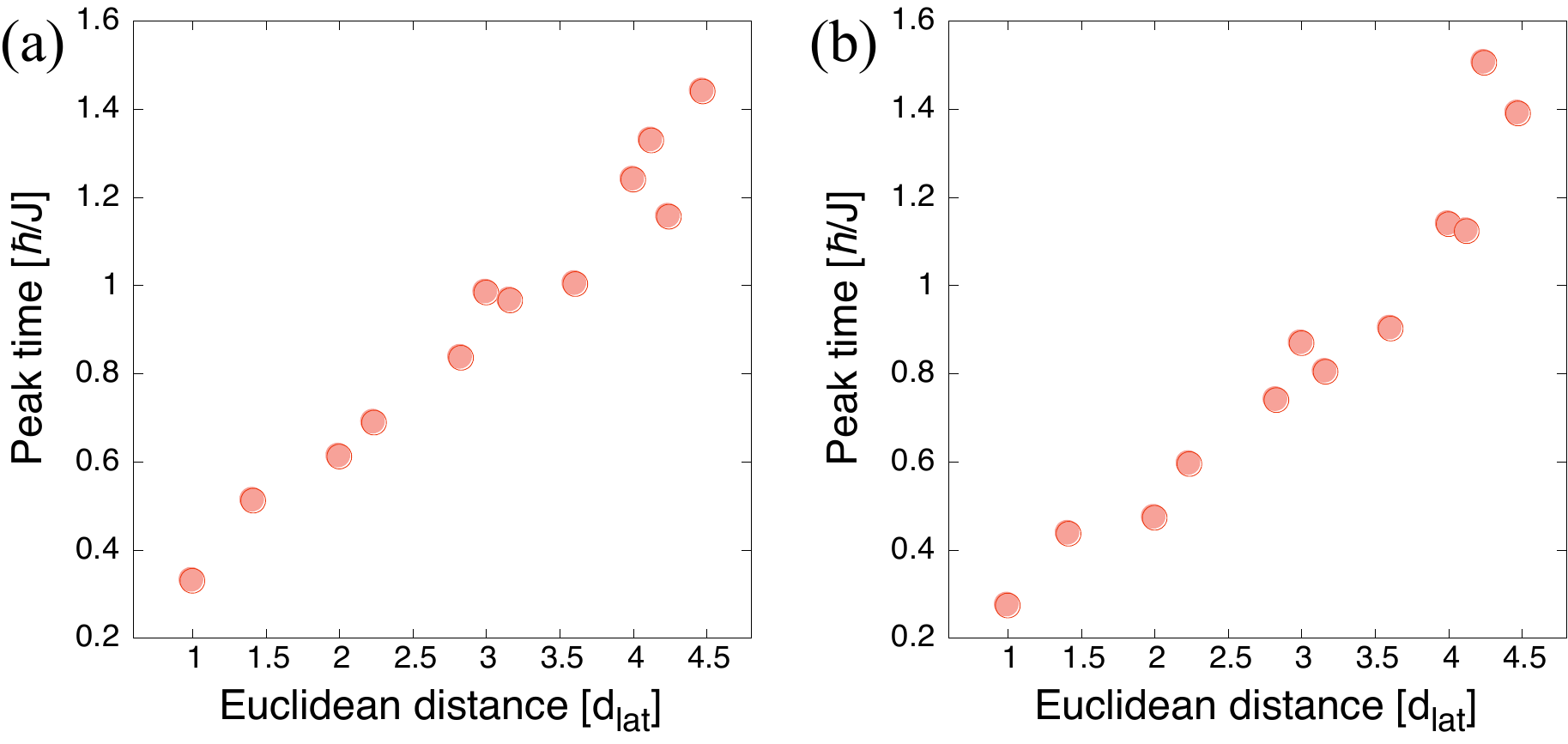}
\vspace{-0.5mm}
\caption{Extracted peak times from the correlation signals at (a) $\lambda_{\rm f}=2$ and (b) $\lambda_{\rm f}=4$ quenched from the Mott-insulator state. The vertical and horizontal axes indicate the peak time and Euclidean distance, respectively.}
\label{figure009}
\end{center}
\end{figure}

\begin{figure}
\begin{center}
\includegraphics[width=70mm]{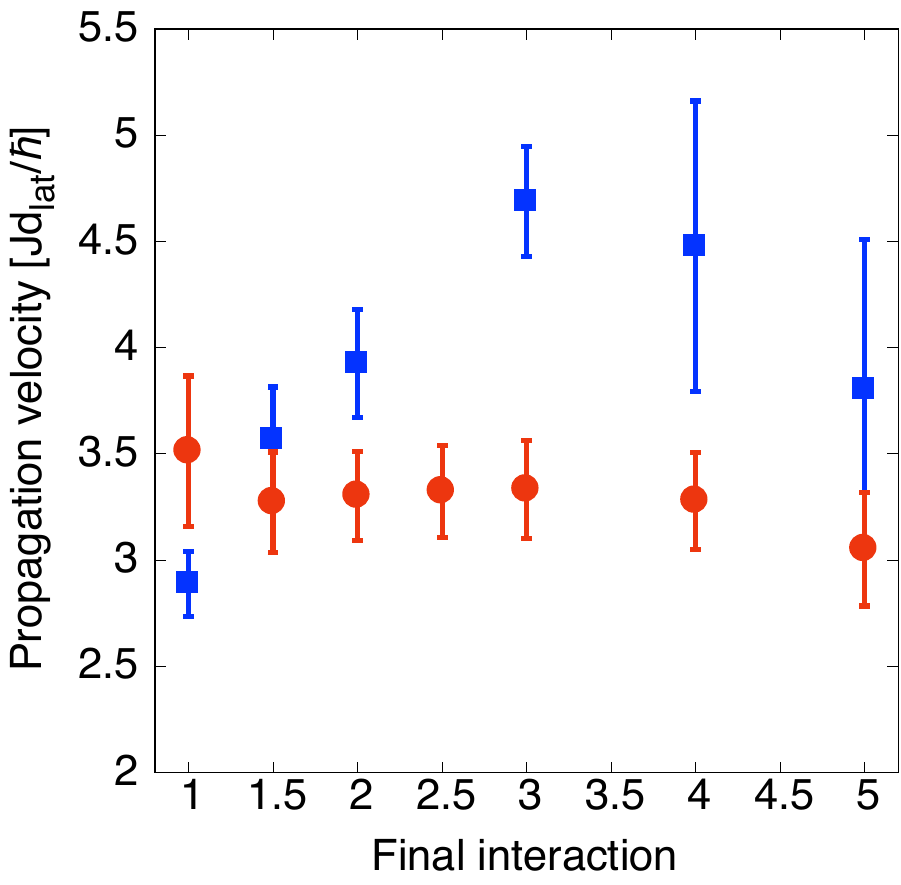}
\vspace{1mm}
\caption{Final interaction dependence of the mean propagation velocity ${\bar v}_{\rm p}$ (red circle). The blue square represents the result of the case of the coherent state shown in Fig.~\ref{figure006}. The vertical bar indicates the normal estimation error of the linear fitting of the peak times.}
\label{figure010}
\end{center}
\end{figure}

Figure \ref{figure008} displays the TWA simulation of the density-density correlation after the sudden quench from the Mott-insulator state with ${\bar n}=10$ at $\lambda_{\rm f}=2$ and $M=20^2$. For the simulation, we utilize the approximate Wigner function (\ref{eq: gauss}). In the result, we can observe a different behavior from the case of the coherent state that a wave packet propagates as a single-peak signal with no fine oscillation in the correlation function. In this case, the velocity of the wave packet can be directly estimated from the activation time of the minimum peak itself. In Fig.~\ref{figure009}, we extract the  peak times from the correlation signals at $\lambda_{\rm f}=2$ and $\lambda_{\rm f}=4$, respectively. In Fig.~\ref{figure010}, we show the propagation velocity ${\bar v}_{\rm p}$ extracted from Fig.~\ref{figure009} as a function of $\lambda_{\rm f}$ and compare it with the results for the coherent state. Figure~\ref{figure010} reveals that ${\bar v}_{\rm p}$ is approximately independent of $\lambda_{\rm f}$ in contrast to the coherent-state case.

\begin{figure}
\begin{center}
\includegraphics[width=80mm]{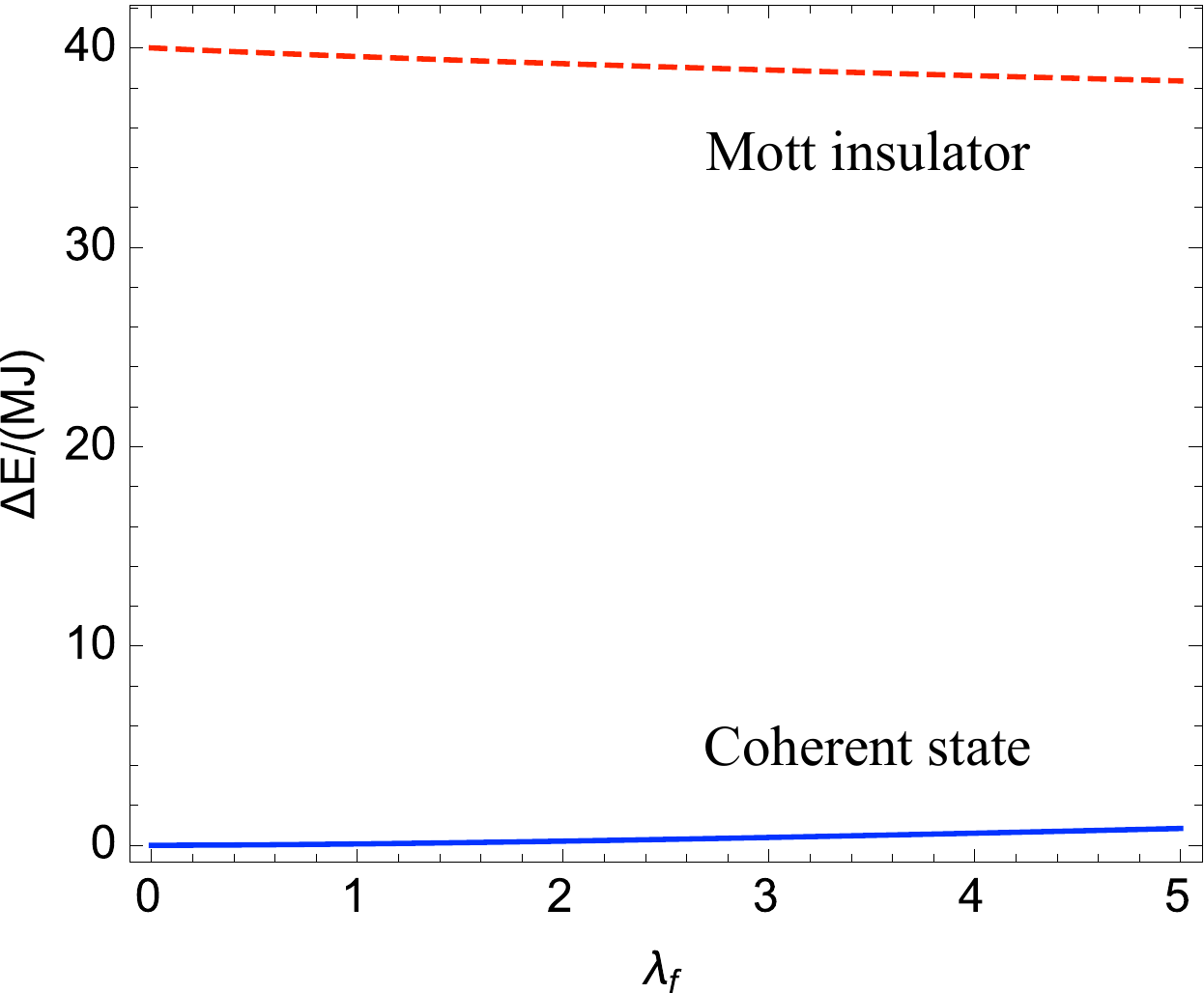}
\vspace{0.5mm}
\caption{Energy deviation $\Delta E/(MJ)$ of the Mott-insulator state from the ground state energy per site of the Hamiltonian at $\lambda = \lambda_{\rm f}$ (red-dashed line). The blue-solid line (same as the one in Fig.~\ref{figure004}) represents the energy deviation when the system is initially prepared in the coherent state.}
\label{figure011}
\end{center}
\end{figure}
 
This qualitative difference can be understood as follows. The Mott insulator state has much larger energy deviation than that of the coherent state as shown in Fig.~\ref{figure011}. This means that the Bogoliubov excitations, which are elementary excitations of the system in the presence of condensates, are no longer relevant to such high-energy dynamics. The sudden quench kicks single-particle excitations with various momenta from the initial density configuration of the Mott insulator state. The absence of the fine oscillation inside the wave packet can be regarded as reflecting an incoherent motion joined by many single-particle excitations. In addition, the single-particle picture can also explain the nearly constant velocity of the correlation spreading. Specifically, within the Hartree--Fock approximation for the Bose particles, the group velocity of the single-particle excitation is independent of $U$ because the interaction effect poses only a constant shift to the non-interacting band \cite{Pethick_2008}. In Appendix~\ref{App: HFA}, we will verify this property by applying the Hartree--Fock approximation to the two-particle Green's function of bosons.  

\section{Conclusions}\label{Sec: conclusions}

In conclusions, we studied the time evolution of the 2D and 3D Bose-Hubbard models after a sudden quench to a weakly interacting regime by using the semiclassical TWA method. We applied the TWA to analyze the redistribution dynamics of the kinetic and onsite-interaction energies after a quench from the singly-occupied Mott insulator state in the 3D Bose-Hubbard model. It was reported that our semiclassical result agrees very well with the experimental one without any fitting parameter. The direct comparison between the TWA and experimental results will be presented in the upcoming paper \cite{Takasu_2018b}.

We also studied the density-density correlation spreading after a sudden quench in the 2D Bose-Hubbard model at a large filling factor. We numerically showed that when the system is initially prepared in the coherent state, then the mean propagation velocity of the correlation wave packet strongly depends on the final interaction strength reflecting the properties of the low-energy elementary excitation in the weakly interacting regime. In contrast, we found that when the initial quantum state is the Mott insulator state, then the mean propagation velocity is almost independent of the final interaction. We also provided a physical interpretation to such a result in terms of the property of the high-energy single-particle excitations.

In a future work, we will develop a similar semiclassical approach making it possible to examine correlation spreading in strongly-correlated regime of the Bose-Hubbard system. For the purpose, we will generalize the SU$(N)$TWA method originally made for interacting spin systems \cite{Davidson_2015,Wurtz_2018}.

\section{Acknowledgement}

We thank Anatoli Polkovnikov for fruitful discussions. We also thank Shimpei Goto for his early collaboration to this work. The authors thank the Yukawa Institute for Theoretical Physics (YITP) at Kyoto University, where this work was initiated during the YITP workshop on ``Quantum Thermodynamics: Thermalization and Fluctuations". This work was supported by KAKENHI from Japan Society for the Promotion of Science (No.~18K03492, No.~18H05228, No.~25220711, No.~17H06138, No.~18H05405), the Impulsing Paradigm Change through Disruptive Technologies (ImPACT) program, CREST from Japan Science and Technology Agency No.~JPMJCR1673, and the Matsuo Foundation. M. K. was supported by Grant-in-Aid for JSPS Research Fellow Grant Number JP16J07240.

\appendix

\section{Hartree--Fock approximation for the two-particle Green's function}\label{App: HFA}

According to Ref.~\cite{Kadanoff_1962}, we apply the Hartree--Fock approximation (HFA) to the two-particle Green's function in the Bose-Hubbard model. We assume that the system has a spatially uniform density: $\langle {\hat n}_{j}(0) \rangle = \langle {\hat n}_{j}(t) \rangle = {\bar n}$. To simplify the following discussion, we deal with the 1D system. The main result of this appendix [Eq.~(\ref{eq: band shift})] is independent of the dimensionality. 

We consider the one- and two-particle Green's function of the lattice bosons,
\begin{align}
&G_{j,j'}(t,t')=\frac{1}{i}\left\langle T\left\{ {\hat a}_{j}(t){\hat a}^{\dagger}_{j'}(t') \right\}\right\rangle, \nonumber \\
&G^{(2)}_{j_1,j_2,j_1',j_2'}(t_1,t_2,t_1',t_2')=  \nonumber \\
&\;\;\;\;\;\;\;\;\;\;\;\;\;\;\;\; \frac{1}{i^2}\left\langle T\left\{ {\hat a}_{j_1}(t_1){\hat a}_{j_2}(t_2){\hat a}^{\dagger}_{j_2'}(t_2'){\hat a}^{\dagger}_{j_1'}(t_1') \right\}\right\rangle, \nonumber
\end{align}
where $T\left\{ \cdots \right\}$ indicates a time-symmetric ordering for operator products inside the bracket. From the Heisenberg equation for ${\hat a}_{j}(t)$, $G_{j,j'}(t,t')$ obeys the following equation of motion: 
\begin{align}
i\hbar\frac{\partial}{\partial t}&G_{j,j'}(t,t') + JG_{j+1,j'}(t,t') + JG_{j-1,j'}(t,t')  \nonumber \\
&-iU \left. G^{(2)}_{j,j,j',j}(t,t_1,t',t_1+\delta)\right|_{t_1=t}= \hbar \delta(t-t')\delta_{j,j'} \nonumber
\end{align}
where $\delta$ is a positive and infinitesimal shift.

In the HFA, $G^{(2)}_{j,j,j',j}(t,t_1,t',t_1+\delta)$ is factorized into two parts as follows:
\begin{align}
G^{(2)}_{j,j,j',j}(t,t_1,t',t_1+\delta) = &\;\;G_{j,j'}(t,t')G_{j,j}(t_1,t_1+\delta)  \nonumber \\
+ &\;\;G_{j,j}(t,t_1+\delta)G_{j,j'}(t_1,t'). \nonumber
\end{align}
This treatment can be regarded as a mean-field approximation, where any correlation between two indistinguishable bosons are neglected \cite{Kadanoff_1962}. At $t_1=t$, we find that 
\begin{align}
G_{j,j}(t,t_1+\delta)=G_{j,j}(t_1,t_1+\delta)=-i\langle {\hat n}_{j}(t) \rangle = -i{\bar n}. \nonumber
\end{align}
Thus, the equation of motion results in a closed equation:
\begin{align}
&\left\{ i\hbar\frac{\partial}{\partial t} - 2U{\bar n} \right\}G_{j,j'}(t,t')  \\
&+ JG_{j+1,j'}(t,t') + JG_{j-1,j'}(t,t') = \hbar \delta(t-t')\delta_{j,j'}.\nonumber
\end{align}
This equation means a constant shift of the pole of the one-particle Green's function as
\begin{align}
\epsilon_{\rm free}(p) \rightarrow \epsilon_{\rm free}(p) + 2U{\bar n}, \label{eq: band shift}
\end{align}
where $\epsilon_{\rm free}(p)=-2J{\rm cos}(pd_{\rm lat}/\hbar)$ is the single-particle dispersion at $U=0$. This result says that the interaction does not change the group velocity of the single-particle excitation within the HFA.


\begin{thebibliography}{99}
\bibitem{Bloch_2012} I. Bloch, J. Dalibard, and S. Nascimb\'{e}ne, Nat. Phys. {\bf 8}, 267 (2012).
\bibitem{Gross_2017} C. Gross and I. Bloch, Science {\bf 357}, 995 (2017).
\bibitem{Hofstetter_2018} W. Hofstetter and T. Qin, J. Phys. B: At. Mol. Opt. Phys. {\bf 51}, 082001 (2018).
\bibitem{Trotzky_2010} S. Trotzky, L. Pollet, F. Gerbier, U. Schnorrberger, I. Bloch, N. V. Prokof'ev, B. Svistunov, and M. Troyer, Nat. Phys. {\bf 6}, 998 (2010).
\bibitem{Endres_2011} M. Endres, M. Cheneau, T. Fukuhara, C. Weitenberg, P. Schau\ss, C. Gross, L. Mazza, M. C. Ba$\tilde{\rm u}$ls, I. Bloch, and S. Kuhr, Science {\bf 334}, 200 (2011).
\bibitem{Trotzky_2012} S. Trotzky, Y-A. Chen, A. Flesch, I. P. McCulloch, U. Schollw\"ock, J. Eisert, and I. Bloch, Nat. Phys. {\bf 8}, 325 (2012).
\bibitem{Cheneau_2012} M. Cheneau, P. Barmettler, D. Poletti, M. Endres, P. Schau\ss, T. Fukuhara, C. Gross, I. Bloch, C. Kollath, and S. Kuhr, Nature {\bf 481}, 484 (2012).
\bibitem{Mazurenko_2017} A. Mazurenko, C. S. Chiu, G. Ji, M. F. Parsons, M. Kan\'asz-Nagy, R. Schmidt, F. Grusdt, E. Demler, D. Greif, and M. Greiner, Nature {\bf 545}, 462 (2017).
\bibitem{Polkovnikov_2011} A. Polkovnikov, K. Sengupta, A. Silva, and M. Vengalattore, Rev. Mod. Phys. {\bf 83}, 863 (2011). 
\bibitem{Eisert_2015} J. Eisert, M. Friesdorf, and C. Gogolin, Nat. Phys. {\bf 11}, 124 (2015).
\bibitem{Greiner_2002b} M. Greiner, O. Mandel, T. W. H\"ansch, and I. Bloch, Nature {\bf 419}, 51 (2002).
\bibitem{Altman_2002} E. Altman and A. Auerbach, Phys. Rev. Lett. {\bf 89}, 250404 (2002).
\bibitem{Tuchman_2006} A. K. Tuchman, C. Orzel, A. Polkovnikov, and M. A. Kasevich, Phys. Rev. A {\bf 74}, 051601(R) (2006).
\bibitem{Kollath_2007} C. Kollath, A. M. L\"auchli, and E. Altman, Phys. Rev. Lett. {\bf 98}, 180601 (2007).
\bibitem{Dziarmaga_2012} J. Dziarmaga, M. Tylutki, and W. H. Zurek, Phys. Rev. B {\bf 86}, 144521 (2012).
\bibitem{Meinert-13} F. Meinert, M. J. Mark, E. Kirilov, K. Lauber, P. Weinmann, A. J. Daley, H.-C. N\"agerl, Phys. Rev. Lett. {\bf 111}, 053003 (2013).
\bibitem{Braun_2015} S. Braun, M. Friesdorf, S. S. Hodgman, M. Schreiber, J. P. Ronzheimer, A. Riera, M. del Rey, I. Bloch, J. Eisert, and U. Schneider, PNAS {\bf 112}, 3641 (2015).
\bibitem{Braun_thesis} S. Braun, {\it Negative temperature and the dynamics of quantum phase transitions}, (Ph. D. Thesis, Ludwig-Maximilians-Universit\"at M\"unchen, 2014).
\bibitem{Asaka_2016} H. Asaka, Y. Takasu, and Y. Takahashi, 2016 Autumn Meeting, The Physical Society of Japan, 15aKJ-6 (2016).
\bibitem{Takasu_2018b} Y. Takasu, T. Yagami, H. Asaka, Y. Fukushima, Y. Takahashi, K. Nagao, S. Goto, and I. Danshita, in preparation. 
\bibitem{Hillery_1984} M. Hillery, R. F. O'Connell, M. O. Scully, and E. P. Wigner, Physics Reports {\bf 106}, 121 (1984).
\bibitem{Steel_1998} M. J. Steel, M. K. Olsen, L. I. Plimak, P. D. Drummond, S. M. Tan, M. J. Collett, D. F. Walls, and R. Graham, Phys. Rev. A {\bf 58}, 4824 (1998).
\bibitem{Blakie_2008} P. B. Blakie, A. S. Bradley, M. J. Davis, R. J. Ballagh, and C. W. Gardiner, Adv. Phys. {\bf 57}, 363 (2008).
\bibitem{Polkovnikov_2010} A. Polkovnikov, Ann. Phys. {\bf 325}, 1790 (2010).
\bibitem{Polkovnikov_2002} A. Polkovnikov, S. Sachdev, and S. M. Girvin, Phys. Rev. A {\bf 66}, 053607 (2002).
\bibitem{Polkovnikov_2003a} A. Polkovnikov, Phys. Rev. A {\bf 68}, 033609 (2003).
\bibitem{Polkovnikov_2003b} A. Polkovnikov, Phys. Rev. A {\bf 68}, 053604 (2003).
\bibitem{Polkovnikov_2004} A. Polkovnikov, D.-W. Wang, Phys. Rev. Lett. {\bf 93}, 070401 (2004).
\bibitem{Mathey_2009} L. Mathey and A. Polkovnikov, Phys. Rev. A {\bf 80}, 041601(R) (2009).
\bibitem{Mathey_2010} L. Mathey and A. Polkovnikov, Phys. Rev. A {\bf 81}, 033605 (2010).
\bibitem{Mathey_2017} L. Mathey, K. J. G\"unter, J. Dalibard, and A. Polkovnikov, Phys. Rev. A {\bf 95}, 053630 (2017).
\bibitem{Landea_2015} I. S. Landea and N. Nessi, Phys. Rev. A {\bf 91}, 063601 (2015).
\bibitem{Fujimoto_2018} K. Fujimoto, R. Hamazaki, and M. Ueda, Phys. Rev. Lett. {\bf 120}, 07302 (2018).
\bibitem{Kunimi_2018} M. Kunimi and I. Danshita, in preparation.
\bibitem{Davidson_2015} S. M. Davidson and A. Polkovnikov, Phys. Rev. Lett. {\bf 114}, 045701 (2015).
\bibitem{Schachenmayer_2015a} J. Schachenmayer, A. Pikovski, and A. M. Rey, Phys. Rev. X {\bf 5}, 011022 (2015).
\bibitem{Schachenmayer_2015b} J. Schachenmayer, A. Pikovski, and A. M. Rey, New. J. Phys. {\bf 17}, 065009 (2015).
\bibitem{Wurtz_2018} J. Wurtz, A. Polkovnikov, and D. Sels, Ann. Phys. {\bf 395}, 341 (2018).
\bibitem{Gardiner_2004} C. W. Gardiner and P. Zoller, {\it  Quantum noise: a handbook of Markovian and non-Markovian quantum stochastic methods with applications to quantum optics}, (Springer, 2004).
\bibitem{Rey_2014} A. M. Rey, A. V. Gorshkov, C. V. Kraus, M. J. Martin, M. Bishof, M. D. Swallows, X. Zhang, C. Benko, J. Ye, N. D. Lemke, and A. D. Ludlow, Ann. Phys. {\bf 340}, 311 (2014).
\bibitem{Kordas_2015} G. Kordas, D. Witthaut, and S. Wimberger, Ann. Phys. (Berlin) {\bf 527}, 619 (2015).
\bibitem{Johnson_2017} A. Johnson, S. S. Szigeti, M. Schemmer, and I. Bouchoule, Phys. Rev. A {\bf 96}, 013623 (2017).
\bibitem{Vicentini_2018} F. Vicentini, F. Minganti, R. Rota, G. Orso, and C. Ciuti, Phys. Rev. A {\bf 97}, 013853 (2018).
\bibitem{Altland_2009} A. Altland, V. Gurarie, T. Kriecherbauer, and A. Polkovnikov, Phys. Rev. A {\bf 79}, 042703 (2009).
\bibitem{Orioli_2017} A. P. Orioli, A. Safavi-Naini, M. L. Wall, and A. M. Rey, Phys. Rev. A {\bf 96}, 033607 (2017).
\bibitem{Raventos_2018} D. Ravent\'os, T. Gra\ss, B. Juli\'a-D\'iaz, and M. Lewenstein, Phys. Rev. A {\bf 97}, 052310 (2018).
\bibitem{Davidson_2017} S. M. Davidson, D. Sels, and A. Polkovnikov, Ann. Phys. {\bf 384}, 128 (2017).
\bibitem{Scaffidi_2017} T. Scaffidi and E. Altman, arXiv:1711.04768.
\bibitem{Schmitt_2018} M. Schmitt, D. Sels, S. Kehrein, and A. Polkovnikov, arXiv:1802.06796.
\bibitem{Lauchli_2008} A. M. L\"auchli and C. Kollath, J. Stat. Mech. P05018 (2008).
\bibitem{Barmettler_2012} P. Barmettler, D. Poletti, M. Cheneau, and C. Kollath, Phys. Rev. A {\bf 85}, 053625 (2012).
\bibitem{Lieb_1972} E. H. Lieb and D. W. Robinson, Commun. Math. Phys. {\bf 28}, 251 (1972).
\bibitem{Carleo_2014} G. Carleo, F. Becca, L. Sanchez-Palencia, S. Sorella, and M. Fabrizio, Phys. Rev. A {\bf 89}, 031602(R) (2014).
\bibitem{Krutitsky_2014} K. V. Krutitsky, P. Navez, F. Queisser, and R. Sch\"utzhold, EPJ Quantum Technol. {\bf 1}, 1 (2014).
\bibitem{Jaksch_1998} D. Jaksch, C. Bruder, J. I. Cirac, C. W. Gardiner, and P. Zoller, Phys. Rev. Lett. {\bf 81}, 3108 (1998).
\bibitem{Fisher_1989} M. P. A. Fisher, P. B. Weichman, G. Grinstein, D. S. Fisher, Phys. Rev. B {\bf 40}, 546 (1989).
\bibitem{Greiner_2002a} M. Greiner, O. Mandel, T. Esslinger, T. W. H\"ansch, and I. Bloch, Nature {\bf 415}, 39 (2002).
\bibitem{Kuhner_2000} T. D. K\"uhner, S. R. White, H. Monien, Phys. Rev. B {\bf 61}, 12474 (2000).
\bibitem{Sansone_2008} B. Capogrosso-Sansone, \c{S}. G. S\"oyler, N. Prokof’ev, and Boris Svistunov, Phys. Rev. A {\bf 77}, 015602 (2008).
\bibitem{Sansone_2007} B. Capogrosso-Sansone, N. V. Prokof'ev, and B. V. Svistunov, Phys. Rev. B {\bf 75}, 134302 (2007).
\bibitem{Takasu_2018a} Y. Takasu, Y. Nakamura, J. Kobayashi, H. Asaka, Y. Fukushima, Y. Takahashi, K. Inaba, and M. Yamashita, arXiv:1808.04599.
\bibitem{Gardiner_2002} C. W. Gardiner, J. R. Anglin, and T. I. A. Fudge, J. Phys. B: At. Mol. Opt. Phys. {\bf 35}, 1555 (2002).
\bibitem{Olsen_2004} M. K. Olsen, A. S. Bradley, and S. B. Cavalcanti, Phys. Rev. A {\bf 70}, 033611 (2004).
\bibitem{Altman_2004} E. Altman, E. Demler, and M. D. Lukin, Phys. Rev. A {\bf 70}, 013603 (2004).
\bibitem{Danshita_2009} I. Danshita and P. Naidon, Phys. Rev. A {\bf 79}, 043601 (2009).
\bibitem{Krause_1986} E. F. Krause, {\it Taxicab Geometry} (Courier Dover Publications, New York, 1986).
\bibitem{Pethick_2008} C. J. Pethick and H. Smith, {\it Bose-Einstein Condensation in Dilute Gases} (Cambridge University Press, Cambridge, UK, 2008).
\bibitem{Kadanoff_1962} L. P. Kadanoff and G. Baym, {\it Quantum Statistical Mechanics} (Benjamin, New York, 1962).
\end{thebibliography}
\end{document}